\documentclass[flegn,usenatbib]{mnras}

\usepackage{newtxtext,newtxmath}
\usepackage{mathptmx}
\usepackage{txfonts}

\usepackage[T1]{fontenc}

\DeclareRobustCommand{\VAN}[3]{#2}
\let\VANthebibliography\thebibliography
\def\thebibliography{\DeclareRobustCommand{\VAN}[3]{##3}\VANthebibliography}

\usepackage{graphicx}	% Including figure files
\usepackage{amsmath}	% Advanced maths commands
\usepackage[normalem]{ulem}

\def\lp{\left(}
\def\rp{\right)}
\def\t{\text}
\def\Nbox{N_{{\t{hit, i}}}}
\def\Ntot{N_{\t{tot, i}}}
\def\Nboxback{N_{{\t{hit, back}}}}
\def\Ntotback{N_{\t{tot, back}}}

\title[Monte Carlo CR Transport]{Cosmic Ray Transport in Mixed Magnetic Fields and their role on the Observed Anisotropies}

\author[Fitz Axen et al.]{
Margot Fitz Axen,$^{1,3}$\thanks{E-mail: margotf@lanl.gov}
Julia Speicher,$^{2}$
Aimee Hungerford$^{3}$
and Chris L. Fryer$^{3}$
\\
$^{1}$Department of Astronomy, The University of Texas at Austin, 2515 Speedway, Stop C1400, Austin, Texas 78712-1205, USA\\
$^{2}$American Astronomical Society, 2000 Florida Ave., NW, Suite 300, Washington, DC 20009-1231, USA\\
$^{3}$Center for Theoretical Astrophysics, Los Alamos National Laboratory, Los Alamos, NM 87544, USA
}

\date{Accepted 2020 November 5. Received 2020 October 29; in original form 2020 May 13}

\pubyear{2020}

\begin{document}
\label{firstpage}
\pagerange{\pageref{firstpage}--\pageref{lastpage}}
\maketitle

\begin{abstract}
There is a growing set of observational data demonstrating that cosmic rays exhibit small-scale anisotropies (5-30$^{\circ}$) with amplitude deviations lying between 0.01-0.1\% that of the average cosmic ray flux.  A broad range of models have been proposed to explain these anisotropies ranging from finite-scale magnetic field structures to dark matter annihilation.  The standard diffusion transport methods used in cosmic ray propagation do not capture the transport physics in a medium with finite-scale or coherent magnetic field structures.  Here we present a Monte Carlo transport method, applying it to a series of finite-scale magnetic field structures to determine the requirements of such fields in explaining the observed cosmic-ray, small-scale anisotropies.
\end{abstract}


%%%%%%%%%%%%%%%%%%%%%%%%%%%%%%%%%%%%%%%%%%%%%%%%%%

%%%%%%%%%%%%%%%%% BODY OF PAPER %%%%%%%%%%%%%%%%%%

\section{Introduction}
\label{sec:intro}

Observations of cosmic rays in the TeV-PeV energy range have demonstrated both small- and large-scale anisotropies.  The large ($>60^{\circ}$) and small-scale anisotropies have been observed by a large number of instruments ~\citep{2005ApJ...626L..29A,2006Sci...314..439A,2007PhRvD..75f2003G,2008PhRvL.101v1101A,2009ApJ...692L.130A,2009ApJ...698.2121A,2010ApJ...711..119A,2010ApJ...712.1100M,2010ApJ...718L.194A,2011ICRC....1....6C,2011ApJ...740...16A,2013ApJ...765...55A}. The large-scale dipole anisotropy is reasonably well fit by the asymmetries in nearby sources smoothed by the subsequent diffusion of these cosmic rays~\citep{2006APh....25..183E,2012JCAP...01..011B,2013ApJ...766....4P,2013APh....50...33S}.  It is possible that magnetic fields in the heliosphere could affect the anisotropy~\citep{2013ApJ...762...44D,2014Sci...343..988S}.

The small-scale anisotropies are much more difficult to explain.  
These small-scale anisotropies typically have size scales between $5-30^{\circ}$ and amplitudes between 0.01-0.1\% of the background cosmic ray flux~\citep{2012JCAP...01..011B}.  A number of models have been proposed to explain such anisotropies.   For example, anisotropies could arise from a nearby, as yet undetected, supernova remnant~\citep{2008A&A...485..527S}, perhaps mediated by a local, coherent magnetic field or asymmetry in the propagation~\citep{2008APh....29..420D,2010ApJ...721..750M,2013ApJ...768..124B}. Another set of proposals argue that properties of the heliosphere can drive the observed anisotropies~\citep{2008APh....29..420D,2010ApJ...722..188L,2013ApJ...762...44D}.  More exotic models have also been proposed invoking strangelet production or dark matter annihilation~\citep{2013arXiv1307.6537H,2013PhLB..725..196K}.  More recently, a series of models have proposed that turbulent magnetic fields are sufficient to explain the anisotropies~\citep{2012PhRvL.109g1101G,2014PhRvL.112b1101A}.

Magnetic fields generated in turbulence are believed to exist on all scales, from the smallest scales in the Kolmogorov spectrum to the largest eddy scales\citep{2008RPPh...71d6901K,2012PhRvD..86j3010S,2013JPlPh..79.1011K,2017NJPh...19f5003K,2020ApJ...898...35K}.  In the Milky Way, this leads to magnetic field structures ranging from well below the parsec scale up to a kiloparsec\citep{2008RPPh...71d6901K,2012PhRvD..86j3010S,2013JPlPh..79.1011K}.  If the scale of the magnetic fields were limited to the smallest turbulence scales, particle transport within these fields could be treated in the diffusion limit where the magnetic fields are treated as a scattering term with a net energy loss as is done in codes like GalProp \citep{2007ARNPS..57..285S}. This treatment does not accurately model any possible larger scale, coherent structures in the magnetic field.

However, transport methods that can model both small and large scale fields face numerical challenges.  For instance, the method described in \cite{2007ApJ...659..389F} and \cite{2016ApJ...822..102H} allows for transport in one of two extreme solutions for each spatial zone:  either dominated by small scales (isotropic scattering limit) or dominated by coherent fields. In this paper, we generalize the Monte Carlo method from~\cite{2007ApJ...659..389F} and \cite{2016ApJ...822..102H} to allow for more general magnetic field profiles that include both small- and large-scale features. As in \cite{2016ApJ...822..102H}, we focus on a magnetic field configuration that studies the interaction of a single point source for cosmic ray production with a coherent magnetic field of varying strength relative to the small-scale field (sufficiently small with respect to the spatial scale that the interaction can be treated as isotropic). With this study, we hope to determine the magnetic field properties needed to explain the observed anisotropies in the cosmic ray flux more realistically. 

In section \ref{sec:code}, we describe the properties of our grid, particles and magnetic field configurations. In section \ref{sec:prop}, we describe the methods we use for particle propagation and their differences from previous work. In section \ref{sec:ver} we describe the verification tests we used on the code.  In section \ref{sec:results}, we study the propagation of cosmic rays in a box to apply this code to a cosmic-ray transport application.  In section \ref{sec:observational}, we study the observational implications of these transport calculations. Finally, in section \ref{sec:conc}, we conclude with a discussion of the properties needed to produce cosmic ray anisotropies.

\section{Code Description and Initial Conditions}
\label{sec:code}

\subsection{Grid Geometry and Particle Properties}
\label{sec:grid}

The setup of our grid is similar to that used by \cite{2016ApJ...822..102H}. We define a three dimensional grid of cubic zones, with physical properties that are constant within each zone. The grid we use is a 150x150x150 pc grid, divided into 50 zones of 3x3x3 pc. Though the zone size is arbitrary, it is set to be much larger than the cosmic ray scattering length and Larmor radius of particle motion. We define the tally plane as being the upper x face of the grid, at x=150 pc. The particles are propagated from a chosen starting location through this grid until they hit one of the grid faces or until they are absorbed into the ISM.

The current code physics is currently limited to the propagation of protons, which are expected to be the dominant form of cosmic rays that create the observed anisotropies in the TeV-PeV energy range \citep{2016ApJ...822..102H}.  We assume a point source for cosmic rays placed in the center of the simulation grid, emitting particles in random directions according to an isotropic distribution.  All particles are assumed to stay traveling at relativistic velocities (v $\approx$ c).  We tested primarily particles with energy of 10 TeV, which is at the lower end of the energies observed for cosmic ray anisotropies. Additionally, we did one study in which we varied the cosmic ray energy to higher energy values, up to 100 PeV.  Note that the proton mean free path is what sets the fundamental scale of the computed solutions and there is a straightforward relationship between proton energy and proton mean free path shown in equation~(\ref{eq:ksc}).  Variation of the assumed proton energy has been used as our primary means to study the impact of scaling the proton mean free path. In principle, these particles lose energy due to Coulomb scattering, ionization, and other processes, as was studied in \cite{2016ApJ...822..102H}. However, for the size-scale and conditions in their, and our simulation grid, they found that energy losses of protons were minimal (\textasciitilde 1-10 MeV over the course of their entire propagation) and, for the calculations in this paper, we do not include energy losses.

\subsection{Magnetic Fields}

The code is designed so that the properties of the magnetic field can be varied in each zone. This includes the small-scale magnetic field amplitude and the coherent magnetic field amplitude and direction. For the calculations in this paper, we focus on a bimodal distribution of field structures:  small-scale fields with size-scales well below the zone size, and coherent magnetic fields that are equal to or greater than the zone size. In the extreme case of no coherent magnetic field, we can transport using a statistically sampled isotropic magnetic field direction equivalent to a symmetric diffusion coefficient. Focusing on a single coherent field allows us to test the ability of the code to model small- and large-scale magnetic field structures combined in a single zone and to study in detail the effects of these global structures on the transport. This setup is a considerable improvement over the grid design used by \cite{2016ApJ...822..102H}, whose setup only allowed for either a turbulent field or a coherent field in each zone of the grid. 

For our simulations, we assume that the entire grid is filled with a small-scale magnetic field component of constant amplitude $B_\text{t}=3 \mu$G.  This field is assumed to be produced in the smallest turbulence scales and is much smaller than our simulation grid-scale.  With our grid size-scales, a typical cosmic ray undergoes many pitch-angle scatterings as it transports through the grid.  The time for energy evolution is much longer than the scattering timescale for our high-energy protons~\citep{1998ApJ...492..352S}.  As discussed above, this long energy-evolution timescale and the relatively small simulation grid-size means that, although the proton undergoes many "scatterings", its energy does not vary considerably as the particle transports through the grid~\citep{2016ApJ...822..102H}. Under these conditions, we can mimic the cosmic ray interaction with the small-scale fields as isotropic and assume the energy is constant during this period. Therefore, we can model the turbulent magnetic field through a random vector sampled at every particle step through the simulation:
\begin{equation} \label{eq:bt1} \text{cos}(\theta) = 2\chi_{1} - 1, \end{equation}
\begin{equation} \label{eq:bt2} \phi = 2\pi \chi_{2}, \end{equation}
where $\chi_i$ are standard deviates between 0 and 1. Converting these expressions to Cartesian coordinates gives a three-component, random vector.  

\begin{figure*}
\centering
\includegraphics[width=0.45\linewidth]{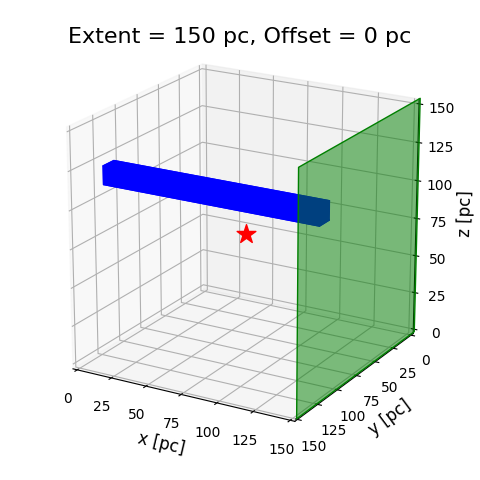}
\includegraphics[width=0.45\linewidth]{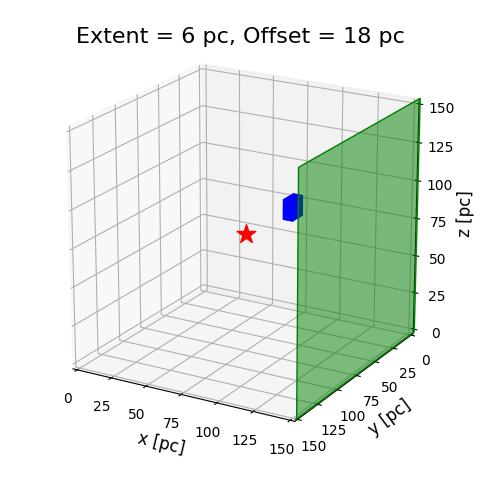}
\includegraphics[width=0.45\linewidth]{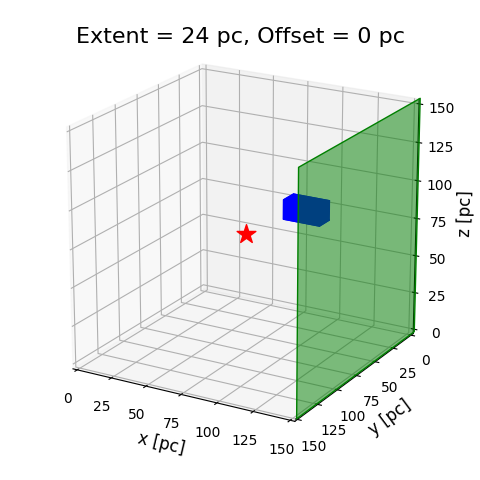}
\includegraphics[width=0.45\linewidth]{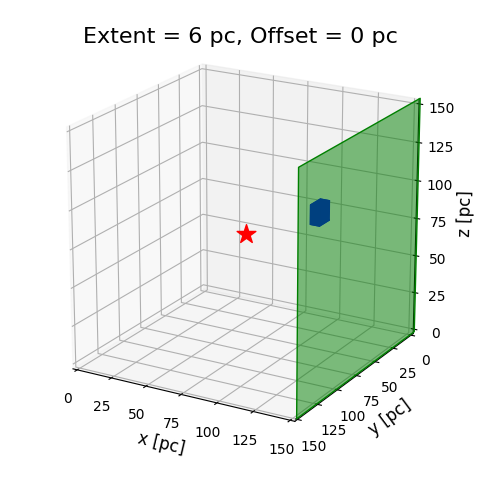}
\caption{Comparison of a few of our simulation setups to that used by ~\protect\cite{2016ApJ...822..102H}. The source of cosmic rays is placed in the center of our simulation grid, shown by the red star. Along a rectangular path, shown by the blue box, we place a coherent magnetic field on top of the small-scale field.  The amplitude of this coherent magnetic field is varied as a fraction of the amplitude of the small-scale field. We vary the length of the coherent magnetic field and its offset from the tally plane, shown by the green region. The top left plot shows Harding's setup, running the length of the simulation space, while the other three show some variations of the box offset and extent which we use.}
\label{fig:setup1}
\end{figure*}

In addition to this turbulent field, we define six grid zone boundaries that bound a region that includes a coherent magnetic field component. This single coherent field is only speculative but is meant to approximate some of the features of the broad range of magnetic field scales that may exist in the ISM. Inside this region, the coherent field component has a constant magnitude and direction given by $B_\text{g}(x,y,z)=(B_\text{g}, 0, 0)$. Clearly, modeling the coherent magnetic field in this way is an oversimplification of the net effects large scale magnetic field structures in the ISM would have, but it allows us to test the importance of these possible large-scale coherent fields. We study a variety of different configurations for the box region containing the coherent component to the magnetic field. We have chosen to keep the y and z coordinates of the box zone numbers constant at 108 pc $<$ y,z $<$ 120 pc. The geometry of the coherent field region is varied using the x zone value of the box, for which we vary the offset from the tally plane and the extent of the box. We tested box extents of 6 and 24 pc and varied the box offset from the tally plane from 0 to 18 pc. This is in contrast to \cite{2016ApJ...822..102H}, whose coherent magnetic field region spanned the entire x length of the simulation space. A comparison of this is shown in Figure \ref{fig:setup1}. 

The amplitude of the coherent magnetic field is varied as a fraction of the maximum amplitude of the small scale field. The exact values for the three parameters studied are shown in Table \ref{tab:runs}.

\begin{table} 
\centering
\caption{Shows the full set of test configurations we chose. Altogether there were 42 different coherent magnetic field configurations.}
\label{tab:runs}
\begin{tabular}{ |c|c| } 
\hline
Variable & Tested Values \\
\hline
 Extent [pc] & 6, 24 \\
 Offset [pc] & 0, 3, 6, 9, 12, 15, 18 \\
 $B_\text{g}$/$B_\text{t}$ & 0.1, 0.5, 1.0 \\
 \hline
\end{tabular}
\end{table}

\section{Particle Propagation}
\label{sec:prop}

In this section, we describe the propagation of particles through our code. The methods we use are similar to those used in \cite{2016ApJ...822..102H}. At the beginning of the particle's lifetime, its starting location is determined by the input source location and its initial direction is sampled randomly from an isotropic distribution. It is then propagated through the grid using one of two methods, depending on whether it is in the coherent magnetic field region or not. 

In \cite{2016ApJ...822..102H}, the methods used included a direct transport Monte Carlo method if the particle was in the coherent magnetic field region and, if not, a Discrete Diffusion Monte Carlo (DDMC) method~\citep{2003JCoPh.189..539E}.  Given the simplicity of the magnetic field configuration, such an approach is reasonable. This DDMC approach reproduces the results of simple diffusion transport schemes such as GalProp~\cite{2007ARNPS..57..285S}. While we were able to use a DDMC algorithm very similar to that used by \cite{2016ApJ...822..102H} for the majority of our grid, the direct transport method did not allow for the magnetic field configurations that include the isotropic, turbulent field component in addition to the coherent field component inside the box. 

To study these hybrid regions, we modified the transport method used by \cite{2016ApJ...822..102H} to account for particle propagation inside the coherent field region.  We tested two different methods for this hybrid transport/diffusion Monte Carlo algorithm. Outside the coherent field region, the particles were propagated using a DDMC method similar to that used by \cite{2016ApJ...822..102H}. In \ref{sssec:trans} we first discuss the direct transport Monte Carlo scheme which was used in \cite{2016ApJ...822..102H} and which is the basis for our hybrid transport method. In \ref{sssec:diff} we discuss the Monte Carlo diffusion approximation which we employ for most of the grid. Finally in \ref{sssec:hybrid} we discuss the two methods we use for our coherent field region in order to account for both components to the magnetic field.

\subsection{Particle Transport Monte Carlo Methods}
\label{sssec:trans}

A particle transport Monte Carlo code tracks the particle motion as the particle goes through a zone and passes into the next. Particles within a magnetic field of strength $B$ propagate according to the equation of motion:
\begin{equation} \label{eq:vel} 
\frac{dv}{dt} = \frac{c}{r_{\text{L}}}v \times \hat{B}, 
\end {equation}
where $\hat{B}$ is the direction of the net magnetic field, $v$ is the direction of the particles velocity, and  $r_{\text{L}}$ is the Larmor radius of the particle motion. This causes particles to move in a circular path of radius $r_{\text{L}}$ around the magnetic field line, perpendicular to it. The component of the particle's initial velocity parallel to the magnetic field line does not change. Particles within a magnetic field of strength $B$, energy $E$, and charge $Ze$ have a Larmor radius of \citep{2002cra..book.....S}: 
\begin{equation} \label{eq:larmor}
r_{\text{L}} = 1.1\times10^{-3} Z^{-1} \lp\frac{E}{\text{TeV}}\rp\lp\frac{B}{\mu \text{G}}\rp^{-1} \text{pc}.
\end{equation}

Solving equation~(\ref{eq:vel}) is computationally difficult due to the number of directional changes the particles make spiraling around the magnetic field lines. Therefore, the approach that \cite{2016ApJ...822..102H} and many others take is to approximate the particle motion as simply following the field line it experiences. Using this approximation, one step through the simulation is a step over which the amplitude and direction of the total magnetic field is constant.  Effectively, the particle's velocity vector is either parallel or antiparallel to the magnetic field line, depending on its initial direction in approaching it. The particle's position can then be reset as:
\begin{equation} \label{eq:transpos}
x(t)= x_0+ \frac{(v_0 \cdot B)}{B}\hat{B}t,
\end{equation}
where $x_0$ is the particle's previous position, $v_0$ is the initial direction of the particle's velocity when approaching the field line, $\hat{B}$ is the magnetic field direction, and $t$ is the time the particle took for the step. The full path length traversed by the particle is $ct$.

Equation~(\ref{eq:transpos}) is assumed to be a valid approximation if the Larmor radius is on par with or smaller than the path length of the particle, under the assumption that solving equation~(\ref{eq:vel}) directly would be unlikely to move the particle into a region with a different magnetic field. We note that it is possible that solving equation~(\ref{eq:vel}) explicitly can alter the particle motion, and should be studied in future work. 

The time $t$ that a particle is expected to take for one step is dependant on the mean free path of the particle motion. By describing the turbulent magnetic field interaction as a scattering term, we can alse describe the distance to scattering interaction. The scattering mean free path describes how far the particle is expected to travel before encountering a change in the turbulent component of the magnetic field, and is determined by the particle energy and the strength of the magnetic field. It is given by \citep{1998ApJ...492..352S, 2007ApJ...659..389F}
 \begin{equation} \label{eq:ksc}
     \lambda_{\text{sc}} = 2\times10^{7}\lp\frac{\lambda_{\text{max}}}{1\text{cm}}\rp^{1/2}\lp\frac{B}{0.1 \text{mG}}\rp^{-1/2} \lp\frac{E}{10 \text{TeV}}\rp^{1/2}\text{cm},
 \end{equation}
 where $\lambda_{\text{max}}$ is the scale length of the turbulent magnetic field, B is the amplitude of the total magnetic field (turbulent and coherent) and E is the proton energy. We use $\lambda_{\text{max}}= 10^{15}$ cm, as was done in \cite{2016ApJ...822..102H}. For the cosmic ray energies we consider, magnetic field scattering is the only significant particle interaction, and other interactions such as proton-proton scattering are negligible. Therefore, we take the total mean free path to be equal to the scattering mean free path: $\lambda=\lambda_{\text{sc}}$.  
 Particles with a mean free path $\lambda$ follow an exponential probability distribution for their motion, $P(x) = e^{-x/\lambda}$, where x is the distance traveled in one step. Solving this distribution for its cumulative distribution function (CDF) gives the expression $\chi=1-e^{-x/\lambda}$, where $\chi$ is a random variable sampled between 0 and 1. This can be rearranged to obtain a distance traveled for that step, along with the time it took for the particle to go that distance: 
\begin{equation} \label{eq:transdis} 
t=x/c= -\text{ln}(\chi) \lp\frac{\lambda}{c}\rp.
\end{equation}

\subsection{Discrete Diffusion Monte Carlo}
\label{sssec:diff}

In regions with only a turbulent magnetic field, we can treat the transport in the diffusion approximation and we use a DDMC method which is very similar to that used by \cite{2016ApJ...822..102H}. In these regions, particle motion changes direction too quickly for it to be computationally feasible to track the particle motion directly. However, particles in an isotropic magnetic field exhibit ``random walk" motion, with their displacement proportional to the square root of their travel time. DDMC methods combine a number of smaller random walk steps that the particle would take into a larger step based on this principle. Rather than picking a distance to collision using equation~(\ref{eq:transdis}), we instead pick a travel time and then sample a particle distance traveled in that time. 

For a three dimensional random walk, the probability of moving distance $R$ after $N$ steps is \citep{lecture_notes_diffusion, 2016ApJ...822..102H} 
\begin{equation} \label{eq:diffprob}
P(R) = 4\pi R^2\lp\frac{3}{2\pi Na^2}\rp^{3/2}\text{exp}\lp\frac{-3R^2}{2Na^2}\rp,
\end{equation} 
where $a$ is the expectation value for displacement in a single step. To approximate the transport motion, we relate the number of steps, $N$, to the total distance traveled by the particle and the average step size: $N=ct/\lambda$. The expected distance traveled in one step $a=\sqrt{\langle x^2\rangle - \langle x\rangle ^2}=\sqrt{\langle x^2\rangle}$ is calculated from the transport equation probability function to be $a=\sqrt{2}\lambda$. Putting in these values gives the expression:
\begin{equation} \label{eq:diffprob2}
P(R)= 4\pi R^2\lp\frac{3}{4\pi ct\lambda}\rp^{3/2}\text{exp}\lp\frac{-3R^2}{4ct\lambda}\rp.
\end{equation}

This expression does not lend itself to sampling directly via an inversion technique, so we instead cast it in a hybrid form using both inversion sampling and rejection sampling techniques. Specifically, we break the full probability distribution $P(R)$ into two components, $g(R)$ and $h(R)$. Defining the constant $C=3/(4ct\lambda)$, it can be seen that:
\begin{equation}
P(R)  \propto Re^{\frac{-2C}{3}R^2} Re^{\frac{-C}{3}R^2} =g(R)h(R).
\end{equation}
The function g(R) is the 2D random walk probability distribution, while h(R) has the correction for the full 3D solution \citep{lecture_notes_diffusion}.

\begin{figure}
\centering
\includegraphics[width = 0.49\textwidth]{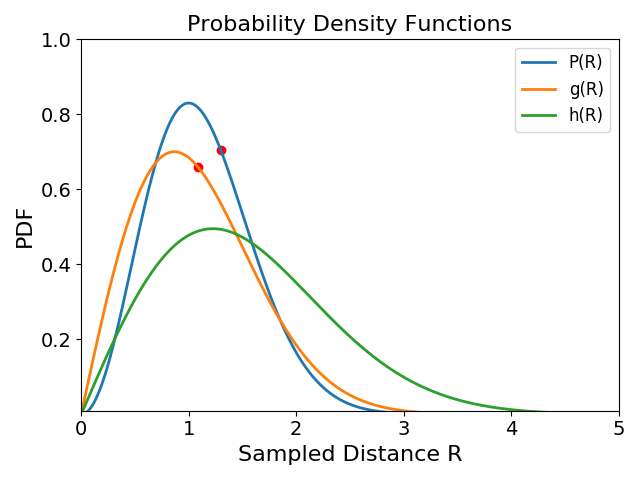}
\caption{Shows the normalized probability density functions using C=1 for the full 3D random walk distribution $P(R)$ (Equation \ref{eq:diffprob2}), the 2D approximation we use $g(R)$ (Equation \ref{eq:diffprob3}), and the 3D correction $h(R)$. The red points show the average value of R sampled for $P(R)$ and $g(R)$.}
\label{fig:pdf}
\end{figure}

To sample the complete function, we first sample a value $R_{\text{samp}}$ via inversion of the CDF for $g(R)$, and then choose to accept $R_{\text{samp}}$ based on a rejection sample of the function $h(R)$. The efficiency of the rejection step is 80$\%$, and does not dramatically increase the computational cost of each run. However, for algorithm simplicity, most of our runs employ an approximate sampling technique where the rejection step is removed.  To justify this, we compared select test cases using the complete function and discovered that, within our error range, there was no noticeable difference in results when the rejection step was removed.  This is largely because $g(R)$ is a close approximation to $P(R)$ in terms of mean diffusion distance properties. The primary effect of removing the rejection check was to slightly decrease the diffusive component's mean free path.  This results in a slight enhancement of the influence of the diffusive field component relative to the coherent field component. As such, using the 2D  approximation to the full 3D solution provides a conservative estimate of the influence of coherent magnetic fields on the cosmic ray flux.
These results can be seen intuitively from Figure \ref{fig:pdf}, which shows the three functions $P(R)$, $g(R)$, and $h(R)$ for C=1. Plotted over $P(R)$ and $g(R)$ in the red points is the average value of $R$ sampled for the two distributions. The function $g(R)$ has a lower peak and is skewed farther towards lower distances than $P(R)$, leading to a slightly lower average value of $R$ sampled.  

With this approximation, the employed probability distribution function is given by:
\begin{equation} \label{eq:diffprob3}
P(R)= Re^{\frac{-2C}{3}R^2}, 
\end{equation}
which yields a sampled distance given by:
\begin{equation}\label{eq:diffdis}
R = \sqrt{-\text{ln}(\chi)2ct \lambda}.
\end{equation}
The values we use for $t$ are $t=1000$ years and $t=2$ years, depending on the particle's situation. We discuss this more in Section \ref{sec:ver}. 

If the particle's travel takes it to the edge of a zone wall before the end of its step, it stops and checks whether it is going into the region with a coherent magnetic field. If it is, it goes to the edge of the zone and exits the diffusion loop, with a travel time of $x/c$. Otherwise, it continues its journey along the path until it either hits another zone wall or finishes its timestep. Therefore all particles will keep going until they either reach the end of the sampled distance or hit the edge of a zone in which the transport mechanism is necessary. For each substep along its sampled distance that it hits a zone wall, the particle's remaining time along its current path is recalculated as the difference between its originally sampled time and its substep time along that path. Once the particle reaches the end of its timestep, its direction is resampled. 

Note that, although equations~(\ref{eq:bt1}) and (\ref{eq:bt2}) are used in both a transport method and DDMC method to approximate the turbulent magnetic field, they have a different meaning for the two. Because the transport algorithm approximates the particle motion directly, the randomly sampled direction represents the direction of the turbulent magnetic field influence for that particle step. In the diffusion algorithm, this vector represents a direction of particle travel over many steps, through many different magnetic field lines.

\subsection{Modified Transport Monte Carlo: Two Methods}
\label{sssec:hybrid}

For the region inside the coherent magnetic field, there is no clear analytic solution to describe the particle motion, arising from the fact that we can neither assume the diffusion approximation or assume the particle flows along field lines. Instead, the solution lies somewhere between these two extremes.  For the particle transport Monte Carlo method, the magnetic field direction is sampled, and the next direction of particle motion is based on the particle's initial velocity direction with respect to this vector. In contrast, for the diffusion approximation, the particles direction change after many steps is sampled; hence the particle's previous direction of travel doesn't matter. Therefore, for the region containing the coherent magnetic field, we explored two methods for modeling the particle motion which attempt to ``combine" the transport and diffusive methods in different ways. Both of these methods give solutions for particle transport that are somewhere between the isotropic magnetic field solution and the transport solution.

The first method we explored (``Method 1") forces particles to follow either transport motion or diffusive motion inside the coherent magnetic field box. For every step, we sample a random number $r$ between 0 and 1 and compare that to the coherent field amplitude ratio ($g=B_{\text{g}}/(B_{\text{g}}+B_{\text{t}})$). If $r<g$, then the particle uses transport motion and follows the coherent magnetic field with one of two directions, based on the velocity of its previous step. If $r>g$, then the travel direction is sampled randomly. Thus, for example, if the amplitudes of the coherent and turbulent field are the same ($B_{\text{g}}/B_{\text{t}}=1.0$), then $g=0.5$, and a particle that travels inside the box will follow transport motion about half of the time and random walk motion the other half. 

The second method we explored (``Method 2") uses a direction of travel which is the vector sum of the coherent magnetic field vector and the turbulent magnetic field vector. The turbulent magnetic field vector is randomly sampled. The coherent field vector, however, in this case, represents the direction the field line would give the particle rather than the field line direction itself. It is one of two directions; either parallel to the field or antiparallel to it, based on the particle's previous direction of travel. 

A consequence of using a timestep $t=1000$ years outside the coherent magnetic field box is that it was inadequate to simply use the transport solution inside the coherent field box and the diffusion approximation outside. The boundary conditions of the transport solution require that when particles leave the transport box they do not resample their direction; which means they are biased to move away from the box. The distance which particles have to travel to recover random walk motion can be relatively large in our case, since with a timestep $t=1000$ years the average distance a particle travels in one diffusion step is $R(t)=0.78$ pc. This causes a deficit of particles inside the coherent magnetic field region. 

In order to rectify this situation, we made two modifications to stop the particles that entered the coherent field region from quickly propagating away from it. First, we put a ``sheath" of one grid zone around the coherent magnetic field in which the particles followed the transport solution, if they had just exited the field region. Second, once the particles had exited the sheath region into the diffusion regime, they used a timestep of $t=2$ years rather than $t=1000$ years. These modifications allowed us to get the correct precision for particles that entered the field region while running the particles that never entered the field region quickly. 

\section{Code Verification}
\label{sec:ver}

\subsection{Isotropic Magnetic Field Travel Time}

One testable property of an isotropic magnetic field is that the particle motion follows a certain distance distribution. Because the mean free path of the particle motion is the same everywhere, the particles should always follow a root-mean square (rms) distance distribution for their travel time following equation~(\ref{eq:diffdis}) of $R(t)=\sqrt{2ct\lambda}$. To confirm our code reproduces this property, we extended the simulation box to calculate a broad particle distribution.

In order to make this simulation run faster with a larger transport box, we set the mean free path in every grid zone to $\lambda = 0.01 \text{pc}$, which is approximately the mean free path of a 100 TeV- 1 PeV proton in a 3$\mu$G magnetic field (our run simulations, with 10 TeV protons, have a mean free path of approximately $\lambda = 0.001 \text{pc}$. We specified a ``box region" given by 42 pc $<$ y,z $<$ 102 pc and 90 pc $<$ x $<$ 150 pc, in which the packets must follow the transport solution; just as would be done if there was a coherent magnetic field in this region. With this setup, about 65 percent of the total number of particles run entered this region at some point during their lifetime, and they spent on average 12 percent of their lifetime in the box region. 

\begin{figure*}
\centering
\includegraphics[width =0.49\linewidth]{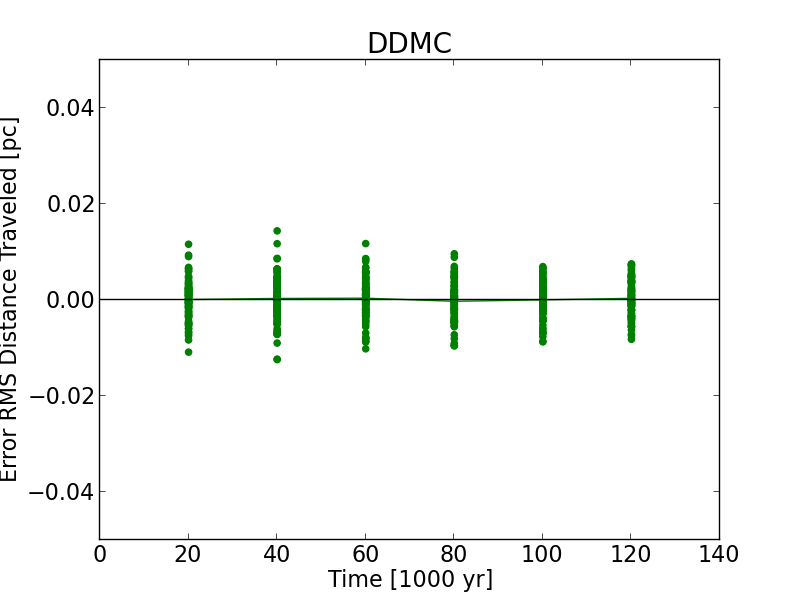}
\includegraphics[width =0.49\linewidth]{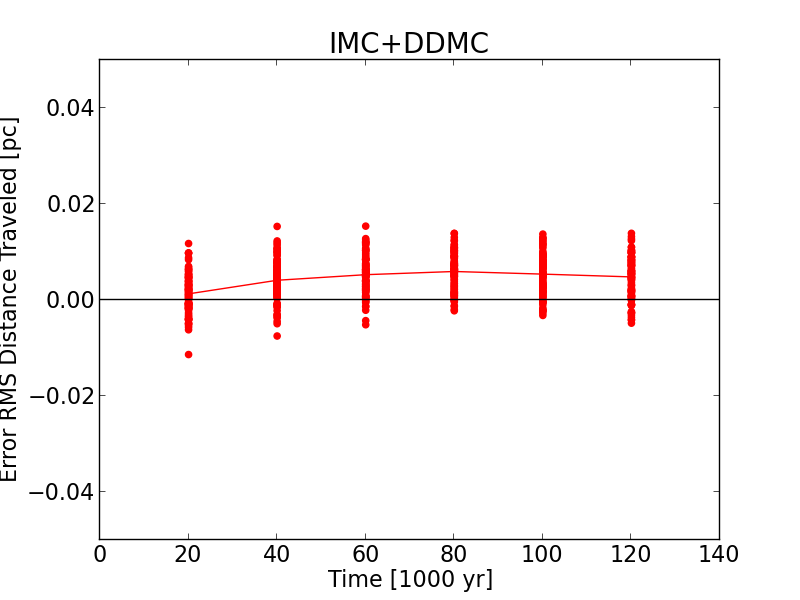}
\caption{Shows, for the isotropic magnetic field  verification, the rms distance errors from the analytic solution calculated at each time of the 100 runs done. The plot on the left uses just the DDMC algorithm, while the plot on the right uses the hybrid DDMC/transport algorithm. The line shows the average of these points at each time. The average error never reaches 1 percent and tends to level out around 80000 years. The distribution of points at every time never spans more than 4 percent.}
\label{fig:rms_times_error}
\end{figure*}

In total, we did 100 runs of the code, each with 10000 particles. Each particle's position was recorded every 20000 years. Figure \ref{fig:rms_times_error} shows the difference between the rms distance traversed for the 10000 particles of a run at each time and the analytic solution at that time. Each point on the plot represents a separate run. The line represents the average of these 100 points at each time. This test was done for both just DDMC and the IMC/DDMC hybrid.

As can be seen, the points for the IMC/DDMC hybrid method does not reproduce the exact solution (averaging above 0). However, the average is below 1\%, and does not show a strongly increasing trend with time. The distribution of the points never exceeds 4 percent.  For this paper, this error lies below the numerical uncertainty in our Monte Carlo statistics (Section~\ref{sssec:stat}) and the errors from our methods are always less than these statistical errors. 

\subsection{Statistical Uncertainties}
\label{sssec:stat}

The goal of this paper is to study the effect of a coherent magnetic field on the particle flux at the tally plane. Although it alters the flux across this entire boundary, the primary affect is at the position of the coherent magnetic field box (in the space 108 pc $<$ y,z $<$ 120 pc).  With a finite set of particles, statistical uncertainties often dominate the errors in a Monte Carlo approach.  To test this, we first ran the particles through a purely isotropic magnetic field, using the diffusion algorithm alone. We then defined the particle flux $F$ for a magnetic field model $i$ as being the ratio between the number of particles that hit this region $\Nbox$ to the number that hit this region in the isotropic run $\Nboxback$, relative to the total numbers of particles $\Ntot$ and $\Ntotback$ run for both:
\begin{equation}
F= \frac{\frac{\Nbox}{\Ntot}} {\frac{\Nboxback}{\Ntotback}}=
\frac{\Ntotback}{\Ntot}\frac{\Nbox}{\Nboxback},
\end{equation}
Statistical errors in Monte Carlo methods cause different runs to give slightly different numbers for this quantity.  However, a set of runs with the same magnetic field configuration should form roughly a Gaussian distribution for the computed value of $F$, with 95$\%$ of the values within $2\sigma$ of the mean. Because for every box configuration we only wanted to do one run, we had to relate the  $1\sigma$ value $\epsilon_{\text{F}}$ of this distribution to an uncertainty measure in a single run based on the number of particle counts. 

For counting statistics, the uncertainty in a counted number of particles is equal to the square root of the number of particles for the statistic. We can therefore compute upper and lower uncertainty bounds for the computed flux above the background using counts for the number of particles that exit the box region and counts that exit the box region for the background run:
\begin{equation} F_{\t{up}} = \frac{\Ntotback}{\Ntot}\frac{\Nbox + \sqrt{\Nbox}}{\Nboxback - \sqrt{\Nboxback}},\end{equation}
\begin{equation} F_{\t{low}} = \frac{\Ntotback}{\Ntot}\frac{\Nbox - \sqrt{\Nbox}}{\Nboxback + \sqrt{\Nboxback}}.\end{equation}
The difference between these two quantities $\Delta F$ gives another measure for the $1\sigma$ uncertainty interval around the computed value for the flux above the background: $ F \pm 1\sigma =F \pm \Delta F/2 = F \pm (F_{\t{up}}- F_{\t{low}})/2$. In order to see whether $\Delta F/2$ computed for one run was approximately equal to what the value of $\epsilon_{\text{F}}$ would be for the distribution, we set up a test configuration used in the full set of runs; a box extent of 24 pc, offset of 0 pc, and $B_{\text{g}}/B_{\text{t}}=1.0$; and did 100 separate particle runs for this same configuration. We did this for both Method 1 and Method 2. 

For both configurations, we computed the mean flux $\mu$ and standard deviation $\sigma$ in the flux for the set of runs. We found that using these quantities, the data accurately fit a Gaussian distribution. We also confirmed that the average computed value of $\Delta F/2$ for the set very closely matched the $1 \sigma$ value $\epsilon_{\text{F}}$ of the Gaussian distribution, and consequently the average computed value of $\Delta F/2F$ for the set was close to the value of $ \sigma/\mu$ from the Gaussian distribution; in fact, for the tests, they were only about 10$\%$ apart from each other. For our production runs we wanted to be within 10$\%$ of the "correct" value of F, with 95$\%$ certainty; so, for every box configuration, we ran enough particles such that the $2 \sigma$ value $2\epsilon_{\text{F}}/ F = \Delta F/F$ was less than 0.1. 

\begin{figure*}
\centering
\includegraphics[width = 0.49\linewidth]{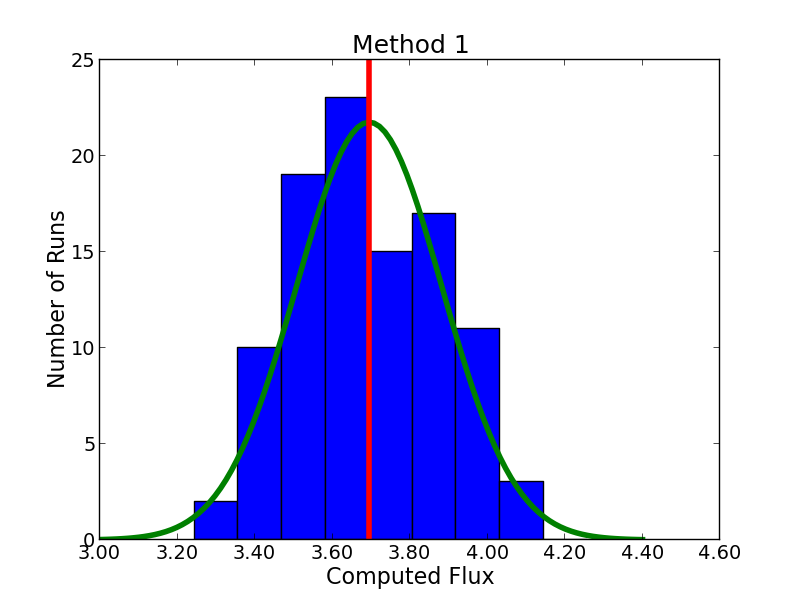}
\includegraphics[width = 0.49\linewidth]{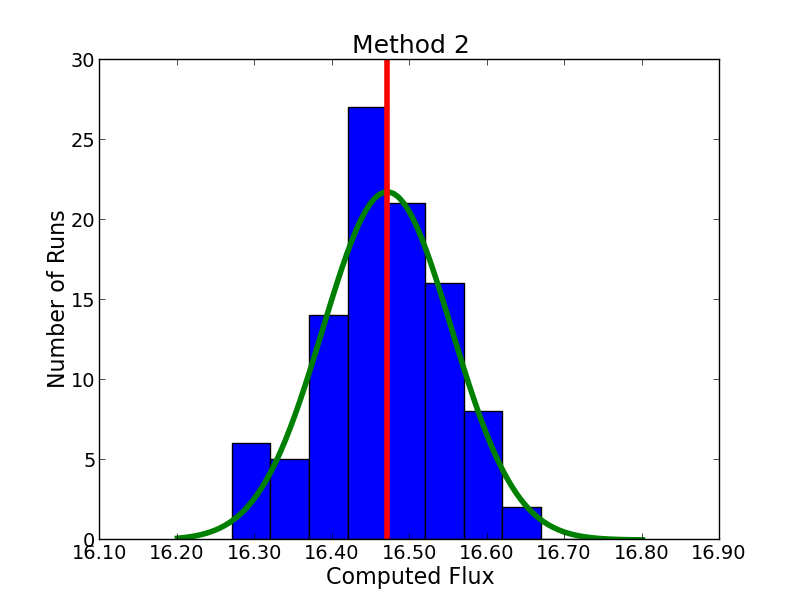}
\caption{Distributions for the uncertainty test described in the text, showing histogram of flux uncertainties obtained for 100 runs of the two methods. The vertical red line in each plot marks the average computed mean of the data and the other tick markers show the 1, 2, and 3 $\sigma$ intervals around the mean. Note: The obtained fluxes are different for the cases due to the different methods used, which is discussed more in Section \ref{sec:results}.}
\label{fig:plot_uncertainty}
\end{figure*}

Figure \ref{fig:plot_uncertainty} shows histograms of the computed flux F above the background for the configuration tested for Method 1 and Method 2 along with a Gaussian fitted to the mean and standard deviations for each. Table \ref{tab:uncertainty} shows the results of the data from the Gaussian distribution. 

\begin{table}
\centering
\caption{Describes the data found by the uncertainty test described in the text. The second column shows the mean value of $\Delta F/2F$ for the set of 100 runs for each configuration (computed for each run individually). The next column shows the ratio between the mean and standard deviation for the Gaussian distribution of all the flux values computed. The fourth column shows the percentage difference between these quantities, and the final column shows percentages within range of the mean for the Gaussian distribution of the flux values.}
\label{tab:uncertainty}
\begin{tabular}{ |c|c|c|c|c|c|c| } 
\hline
M & $\langle\Delta F/2F\rangle$ & $\sigma/\mu$ & $\frac{\langle\Delta F/2F\rangle-(\sigma/\mu)}{(\sigma/\mu)}$ & $1,2,3\sigma$ \\
\hline
 1 & 5.60e-2 & 5.08e-2  & 10.22 & 62,96,100 \\
 2 & 5.60e-3 & 5.05e-3 & 10.91 & 70,95,99 \\
\hline
\end{tabular}
\end{table}

\section{Results}
\label{sec:results}

\begin{figure*}
\centering
\includegraphics[width =0.99\linewidth]{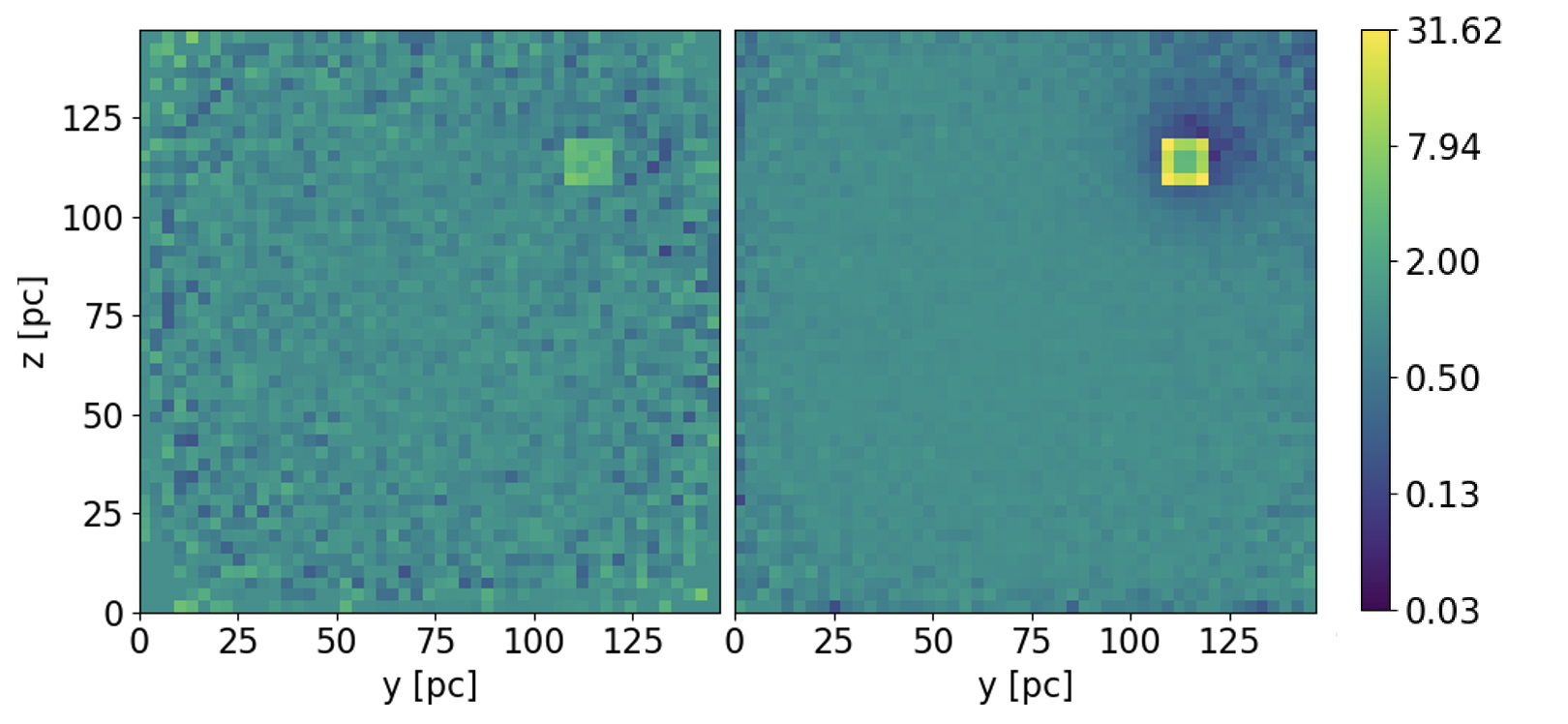}
\caption{Shows two examples of the particle flux above the background $F$ for our simulation runs. Both were done using the same coherent magnetic field configuration of $B_{\text{g}}/B_{\text{t}}=1.0$, box extent=24 pc, and box offset=0 pc, but the left plot was computed using Method 1 while the right plot was computed using Method 2. The coherent magnetic field region can be seen by the yellow and green patches in the upper right corner of the plots.}
\label{fig:es_example}
\end{figure*}

With this code, we can calculate the range of effects varying the coherent magnetic field properties outlined in Section~\ref{sec:grid}. Figure \ref{fig:es_example} shows two contour plots of the particle flux a coherent magnetic field produces above the background at the tally plane. As can be seen, the coherent magnetic field produces a noticeable flux above the background, shown by the yellow and green patches in the upper right corners of the plots. These patches span 4x4 grid zones in our case, and  the flux $F$ previously discussed contains the relative sum of all of the extra particles in this region. 

In these plots, the likelihood of the particles to follow the direction of the coherent field can clearly be observed. One noticeable feature of the plots is that the effect on the tally plane is not always simply a higher flux (darker patch) on the contour plot. Rather, the magnetic field configurations with a stronger effect tend to form a ``ring" of increased flux due to the likelihood of the particles to follow the coherent field when they encounter it in the outer zones of the coherent field region. Another feature of the stronger magnetic field configurations is the decreased flux surrounding the box and extending up into the right corner of the plot. This shadowing effect was also observed in \cite{2016ApJ...822..102H}, and is due to a lack of particles scattering past the coherent field without entering it and getting ``trapped".  Because our coherent field box contains a turbulent component, there is some expectation that the shadowing effect would be muted, but it is still present in many cases. 

\begin{figure*}
\centering
\includegraphics[width =0.49\linewidth]{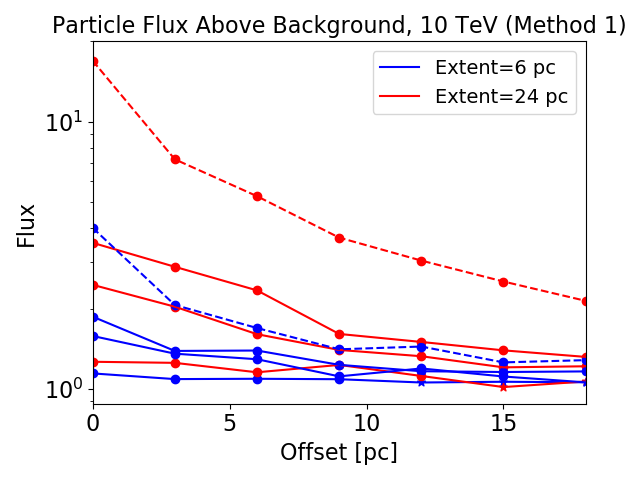}
\includegraphics[width =0.49\linewidth]{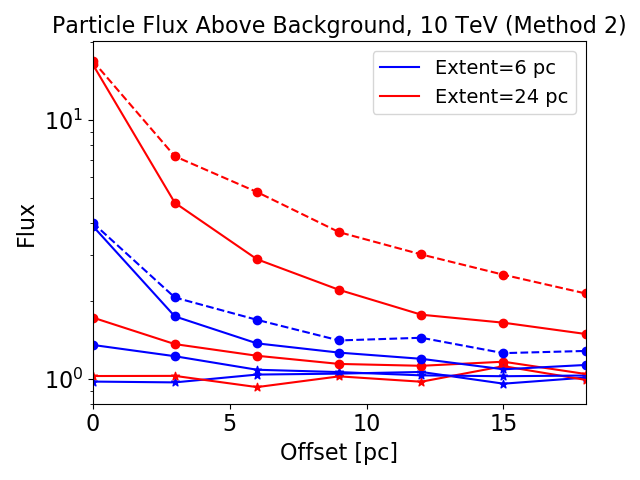}
\caption{Shows the results of our code, with the particle flux above the background defined in Section \ref{sssec:stat} as a function of the coherent field offset from the tally plane. Each color represents a different box extent, which was tested for three different values of the magnetic field. The dotted lines shows values computed using the methods of ~\protect\cite{2016ApJ...822..102H}. Every run is done to 10$\%$ uncertainty. The points show the computed values, with the dots being points whose uncertainty range is above the background while the stars have an uncertainty range below the background.}
\label{fig:results}
\end{figure*}

The computed flux results of our code are shown in Figure \ref{fig:results}. The plot on the left shows the results using ``Method 1" while the plot on the right shows the results using ``Method 2". The figure shows the computed flux above the background F for every box extent, tally plane offset, and coherent magnetic field ratio tested. Each color shows a different coherent magnetic field box extent, with each color having three different lines representing the three different coherent magnetic field ratios. The dotted lines are put in for comparison and were computed using the methods of \cite{2016ApJ...822..102H}; i.e., no turbulent magnetic field in the coherent field box. In general, the topmost line of one color is the highest coherent magnetic field and the coherent field decreases moving down in height; this holds up except for the lower points close to the background of $F=1$, where statistical uncertainty causes more variation.

These results demonstrate the large effect of a coherent magnetic field component on the particle flux above the background. Indeed, one can see from Figure \ref{fig:results} that for all runs done in the study of coherent magnetic field amplitude, coherent field box length, and coherent field box offset from the tally plane, there was a noticeable flux observed above the background level; and above the observed anisotropy level. Only a few of the runs with the lowest flux values did not have a $2 \sigma$ range above $F=1$, which signifies a 95$\%$ chance of a noticeable observed flux. These points are shown by crosses instead of dots in Figure \ref{fig:results}.

\subsection{Method Differences}

Both of the hybrid transport methods, Method 1 and Method 2, produce the same, generally predictable trends for the variation of the tested box configuration variables. Increasing the coherent magnetic field ratio, increasing the box length, and decreasing its distance from the tally plane all increase the observed flux at the plane. However, the two methods do produce quantitatively-different results from each other, depending on the coherent magnetic field ratio used. 

\begin{figure}
\centering
\includegraphics[width = 0.49\textwidth]{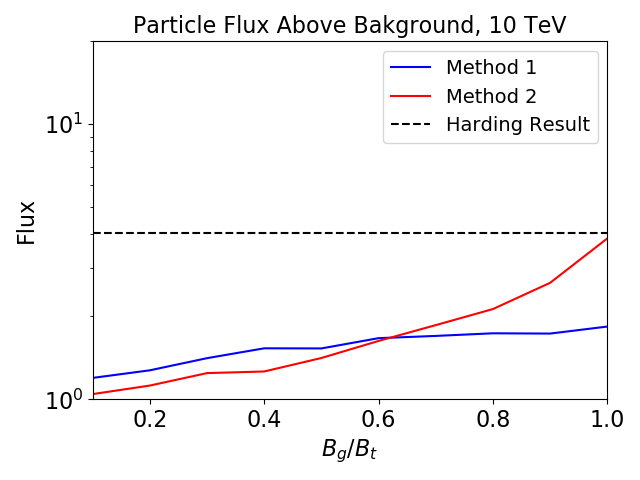}
\caption{Shows the results of varying the magnetic field ratio for one box geometry configuration (extent=6 pc, offset=0 pc).}
\label{fig:resultsoffset}
\end{figure}

In addition to our main configuration parameter space, we did one study where we picked a single box offset of 0 pc and extent value of 6 pc and varied the value of the coherent magnetic field ratio for 10 different values, ranging from 0.1 to 1.0. This is shown in Figure \ref{fig:resultsoffset}. It can be seen that, for higher magnetic field ratios ($B_{\text{g}}/B_{\text{t}}\gtrsim 0.6$), Method 2 predicts a higher particle flux above the background than Method 1 while, for lower magnetic field ratios ($B_{\text{g}}/B_{\text{t}}\lesssim 0.6$), Method 1 predicts a higher particle flux above the background than Method 2. This means that over the span of magnetic field ratio values we tested for the main study, Method 2 has a much broader range in predicted flux values (because higher coherent magnetic field ratios predict a higher particle flux above the background in general). 

\begin{figure*}
\centering
\includegraphics[width = 0.99\linewidth]{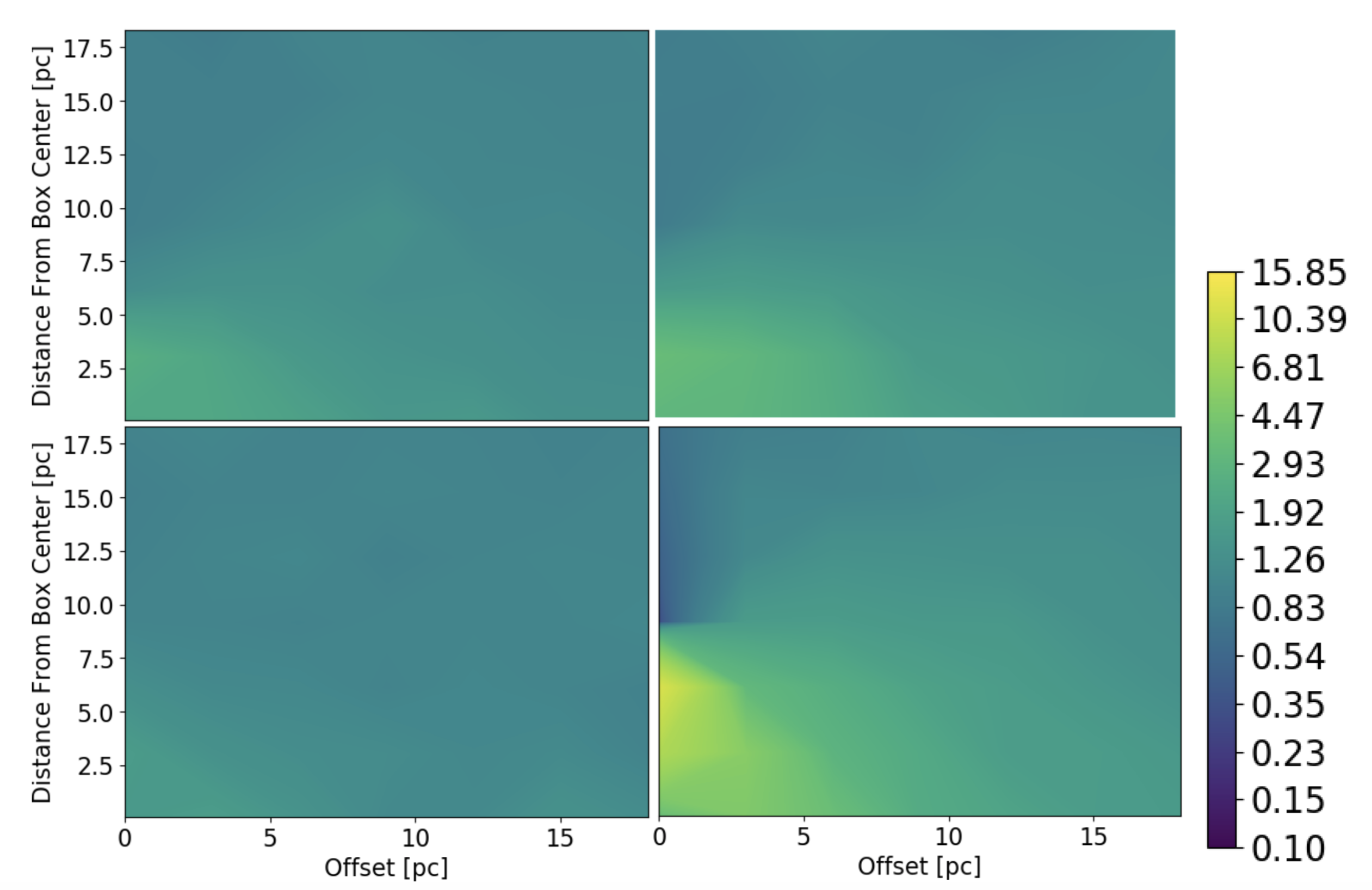}
\caption{Shows the number of particles at the Earth sky as a function of box offset and radial distance from the center of the box region. The top plots were computed using Method 1 and the bottom plots were computed using Method 2. The left column uses a magnetic field ratio of $B_{\text{g}}/B_{\text{t}}=0.5$ while the right column uses a magnetic field ratio of $B_{\text{g}}/B_{\text{t}}=1.0$. Both have a box extent of 24 pc.}
\label{fig:radial}
\end{figure*}

These differences can be seen in the dependence of the results on the spatial distribution of the coherent magnetic field.  Figure \ref{fig:radial} shows the particle flux at tally plane as a function of the distance away from the center of the box region (108 pc $<$ y,z $<$ 120 pc). The top two plots were computed using Method 1, and the bottom two plots were computed using Method 2. The left column uses a coherent magnetic field ratio of $B_{\text{g}}/B_{\text{t}}=0.5$ while the right column uses a coherent magnetic field ratio of $B_{\text{g}}/B_{\text{t}}=1.0$. As can be seen, for the higher magnetic field ratio, Method 2 produces a higher particle flux around the box region, and a greater deficit outside. However, for the lower field value, Method 1 still has an observable flux around the box while it is much harder to see for Method 2.  

The differences between these two methods provide an indication of the uncertainty in the predicted particle anisotropy at the tally plane. For our tests, Method 1 predicts an anisotropy for all box geometries with $B_{\text{g}}/B_{\text{t}}=0.5$ and $B_{\text{g}}/B_{\text{t}}=1.0$, and for most of the geometries with $B_{\text{g}}/B_{\text{t}}=0.1$. Method 2, on the other hand, only predicts an anisotropy for all box configurations with $B_{\text{g}}/B_{\text{t}}=1.0$; all of the configurations with $B_{\text{g}}/B_{\text{t}}=0.1$ and many with $B_{\text{g}}/B_{\text{t}}=0.5$ are not predicted to be high enough to be seen within our 10\% simulation errors. 

\subsection{Variation of Cosmic Ray Energy}

\begin{figure}
\centering
\includegraphics[width = 0.49\textwidth]{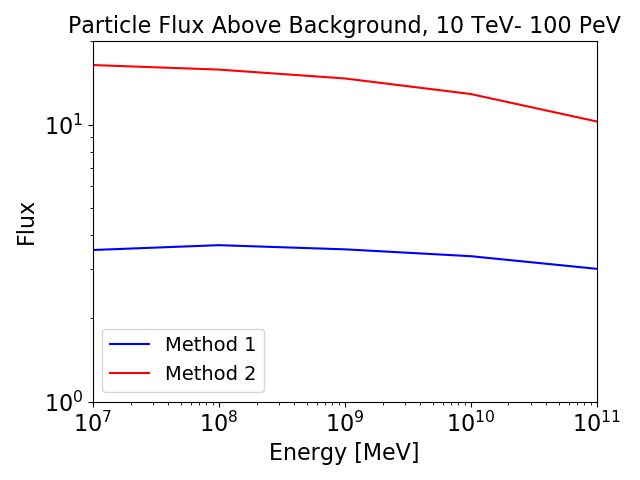}
\caption{Shows the results of varying the energy for one box geometry and magnetic field ratio (extent=6 pc, offset=0 pc, $B_g/B_t=1.0$) }
\label{fig:resultsenergy}
\end{figure}

Finally, we tested one magnetic field configuration using higher energy particles, for both Method 1 and Method 2. We tested 100 TeV, 1 PeV, 10 PeV, and 100 PeV particles. The magnetic field configuration we chose to test was a box extent of 24 pc, offset of 0 pc, and coherent magnetic field ratio of $B_{\text{g}}/B_{\text{t}}=1.0$. We found that the trend is that higher energy particles produce a smaller flux above the background in the box. This is shown in the Figure \ref{fig:resultsenergy}. The reason for this likely has to do with the higher energy particles having a longer path length, which means they are more likely to "escape" from the coherent magnetic field.

\section{Observational Implications}
\label{sec:observational}

We now attempt to approximate the effect that a coherent magnetic field might have on the observed cosmic ray flux at Earth.  Though we have demonstrated theoretically that a coherent magnetic field can create anisotropies in the cosmic ray flux across a tally plane, this doesn't correlate exactly to the anisotropies observed in the cosmic ray arrival direction at Earth. The observed flux at a point requires tallying the angular distribution of the flux as well.  For statistical results, section~\ref{sec:results} tallied the entire flux.  In this section, we also study the angular distribution.

To understand the angular distribution of the material on our tally plan, we ran one configuration where we also tallied the velocity $\vec{u}$ of the particles as they hit the plane. We ran particles using the configuration tested for Method 1 with a box extent=24 pc, offset=3 pc, and $B_{\text{g}}/B_{\text{t}}=1.0$ (though any of the runs with an offset greater than 0 pc could have been used). For every particle that hit the plane, we calculated  $\text{cos}(\theta_u) = \hat{u} \cdot \hat{n}$, where $\hat{n}=[1, 0, 0]$ and $\theta_u$ is the angle between its velocity and the plane surface normal. We then tallied the number of particles between $\text{cos}(\theta_u)$ and  $\text{cos}(\theta_u) + \Delta \text{cos}(\theta_u)$ for values of $\text{cos}(\theta_u)$ between 0 and 1. We used 15 bins in $\text{cos}(\theta_u)$, so $\Delta \text{cos}(\theta_u)= 1/15$.

\begin{figure}
\centering
\includegraphics[width=0.49\textwidth]{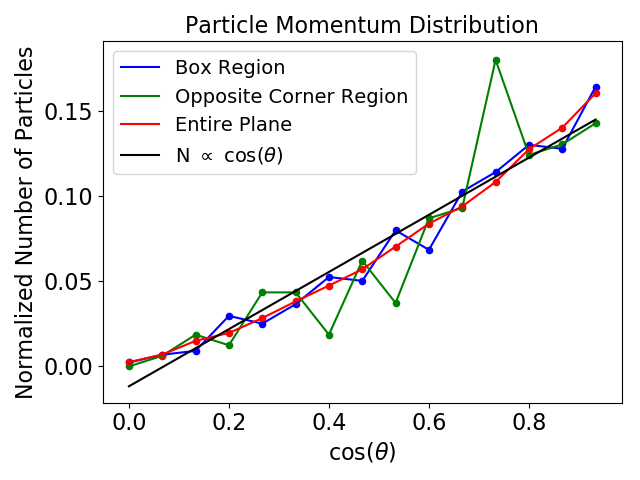}
\caption{ Shows the number of particles emitted in certain different directions as a function of cos($\theta$), run for one box configuration. We show the results for the entire x=150 pc plane (red), as well as the box region alone (blue) and one other region at a different location in the grid (green). Each line is normalized to the number of particles being considered in that region.  The black lines are a linear fit to the red points.  
}
\label{fig:velocity_test}
\end{figure}

We found that the velocity distribution of the particles being emitted from the plane follows the relationship \footnote{This is a Lambertian distribution in angle, which is expected for particles following isotropic scattering motion.}
\begin{equation}
    F(\hat{u}) \propto \hat{u} \cdot \hat{n} = \text{cos}(\theta_u).
    \label{eq:fmu}
\end{equation}
The results of this test are shown in Figure \ref{fig:velocity_test}. Each point represents a total number of particles emitted in one bin. In addition to showing the calculation done for the entire tally plane, we also show the same calculation for two regions on the plane; one that is the box region used to compute the flux, and another region that is in the opposite corner of the plane. All lines are normalized to the number of particles that were considered for that region. These details are to emphasize the fact that although these regions have different total numbers of particles and particle density, they all follow the same distribution in momentum. The main difference between the different regions is the offset of the points from the best fit line; the opposite corner region has a higher offset than the box region or entire plane because there are less particles being considered, leading to higher statistical error. 

We can use the angular distribution of the particles at the tally surface inferred from this study to calculate the observed flux in the observer frame (as a function of viewing angle).  To do so, we determine an observer location ("position of the Earth") at coordinates $[x_e,y_e,z_e]$. We calculate each particle's position in spherical coordinates [$\phi, \theta$] from its Cartesian coordinates $[x_p, y_p, z_p]$ on the tally plane using $\rm tan(\phi)=(y_p-y_e)/(x_p-x_e)$ and $\rm tan(\theta)=(z_p-z_e)/(x_p-x_e)$. Each particle is then in a spherical grid zone $\Omega_z$, where each zone $\Omega_z$ bounds a two dimensional region in [$\phi$, $\theta$] space. By summing only these angle-dependent contributions from the tally surface, we calculate the observed sky distribution for part of the sky that can be mapped from the tally plane. For each spherical grid zone $\Omega_z$, we tallied the proportional flux of particles that would arrive at the Earth point from that zone as 
\begin{equation}
    \label{eq:r_Omega}
   R(\Omega_z) \propto \sum_{i=1}^{N_{\Omega_z}} \rm cos(\theta_{e}),
\end{equation}
where the sum is over all of the $N_{\Omega_z}$ particles that hit the plane in the zone $\Omega_z$ and $\theta_e$ is the angle between the surface normal and the vector connecting the particle hit position to the Earth point. For the plots we present, the Earth point is at position $[x_e,y_e,z_e]=[168, 114, 114]$ pc, and the angular resolution in both dimensions are $\Delta\theta=\pi/50$ and $\Delta\phi=\pi/50$ (50 angular bins each), with $\theta \in [0, \pi]$ and $\phi \in [0, \pi]$. To demonstrate the cosmic ray flux in a way that is commonly done observationally, and reduce the effect of the higher noise floor due to the geometric effect of the angular bins, for each solid angle bin we chose to consider the difference between the flux and the background run flux rather than the ratio:
\begin{equation}
    \label{eq:f_omega}
    F (\Omega_z) = R(\Omega_z)-R_{\t{back}}(\Omega_z) \propto \sum_{i=1}^{N_{\Omega_z}} \t{cos}(\theta_{e})-\sum_{i=1}^{N_{\t{back,}\Omega_z}} \t{cos}(\theta_{e}),
\end{equation}
where here all calculations were done using the same number of particles originally started for both the main and the background runs ($\Ntot=\Ntotback$), since the number of particles will not normalize out in these plots. 

\begin{figure*}
\centering
\includegraphics[width=0.99 \textwidth]{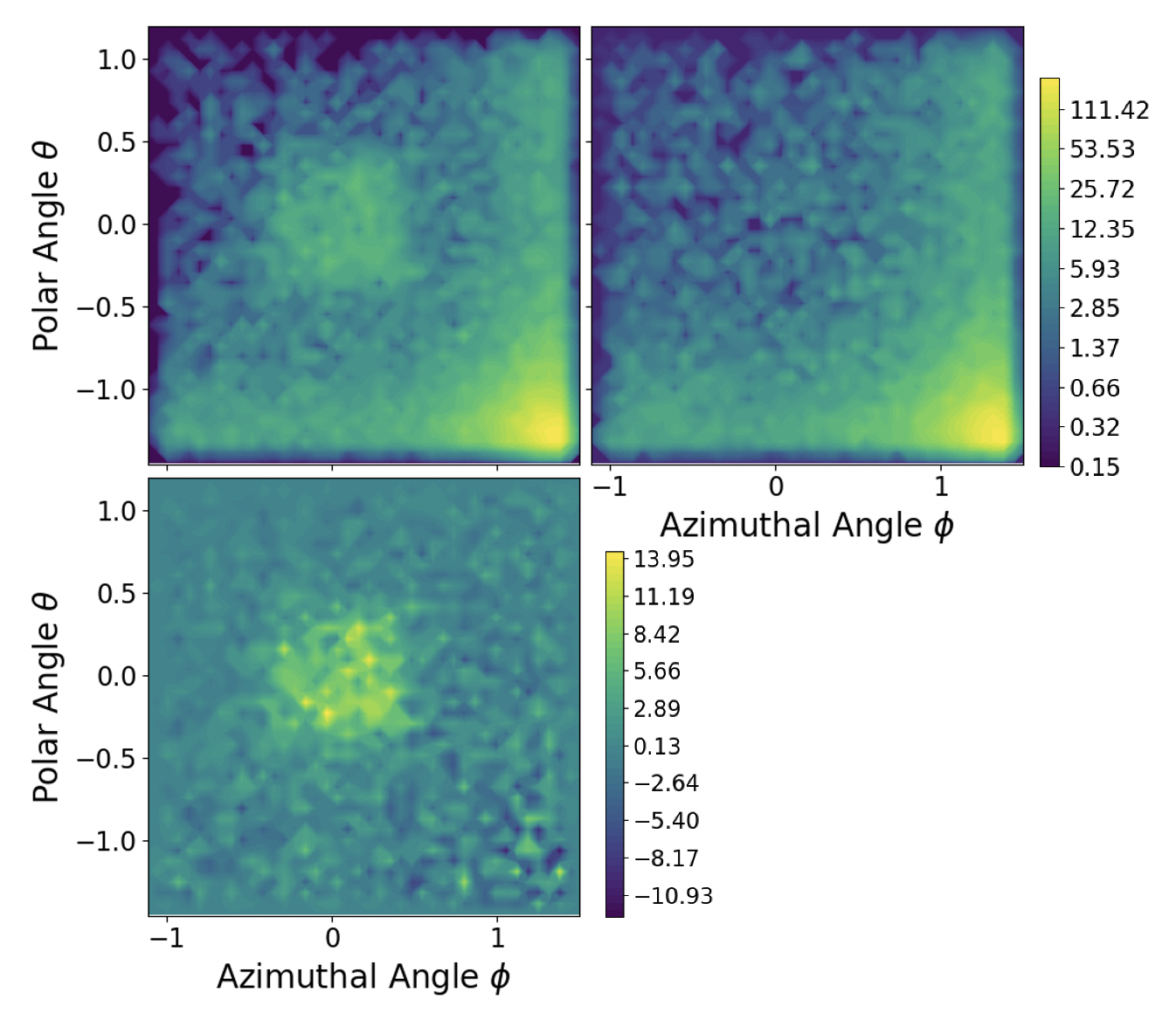}
\caption{Plots of $R(\Omega_z)$ (top left), $R_{\t{back}}(\Omega_z)$ (top right), and $F(\Omega_z)$ (bottom) using Method 1 and $B_{\text{g}}/B_{\text{t}}$=1.0, box extent=24 pc and box offset = 3 pc.}
\label{fig:angle_comparison}
\end{figure*}

Figure \ref{fig:angle_comparison} illustrates this process for one box configuration using Method 2. The top two panels show $R(\Omega_z)$ and $R_{\t{back}}(\Omega_z)$ computed using Equation \ref{eq:r_Omega}. The higher particle counts in the bottom right corners of these panels appear because of the geometric location of the earth sky point compared to the tally plane and the location of the cosmic ray source. The placement of the earth sky point near the top left corner of the tally plane causes more zones on the plane to be included in the lower right angular bins, and fewer in the top left angular bins. Additionally, there are more particle counts in the tally plane zones near the cosmic ray source. In the bottom panel, computed using Equation \ref{eq:f_omega}, these effects are removed and the location of the coherent magnetic field can clearly be seen by the higher particle flux.

\begin{figure*}
\centering
\includegraphics[width=0.99 \textwidth]{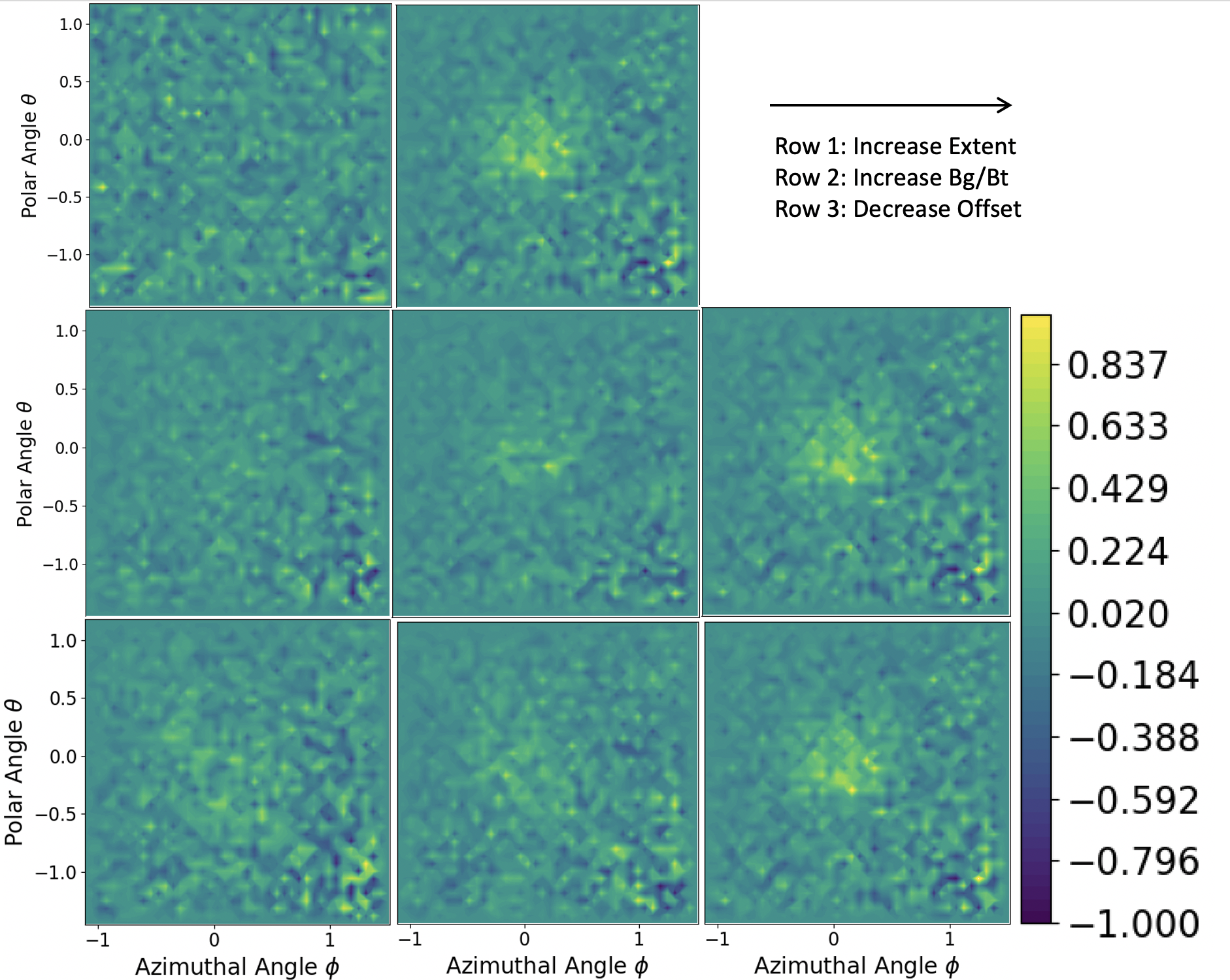}
\caption{Shows plots of $F(\Omega_z)/\sigma_{\Omega}$ for different box configurations using Method 1. The different rows show the effects on the Earth sky image of varying the box extent, coherent magnetic field ratio, and offset. The top row uses box extent values of 6 pc and 24 pc while keeping $B_{\text{g}}/B_{\text{t}}=1.0$ and offset=3 pc. The middle row uses values of $B_{\text{g}}/B_{\text{t}}=0.1$, 0.5, and 1.0 while keeping the box extent=24 pc and box offset=3 pc. The bottom row uses box offset values of 15 pc, 9 pc, and 3 pc while keeping the box extent=24 pc and $B_{\text{g}}/B_{\text{t}}$=1.0.}
\label{fig:method1}
\end{figure*}

\begin{figure*}
\centering
\includegraphics[width=0.99 \textwidth]{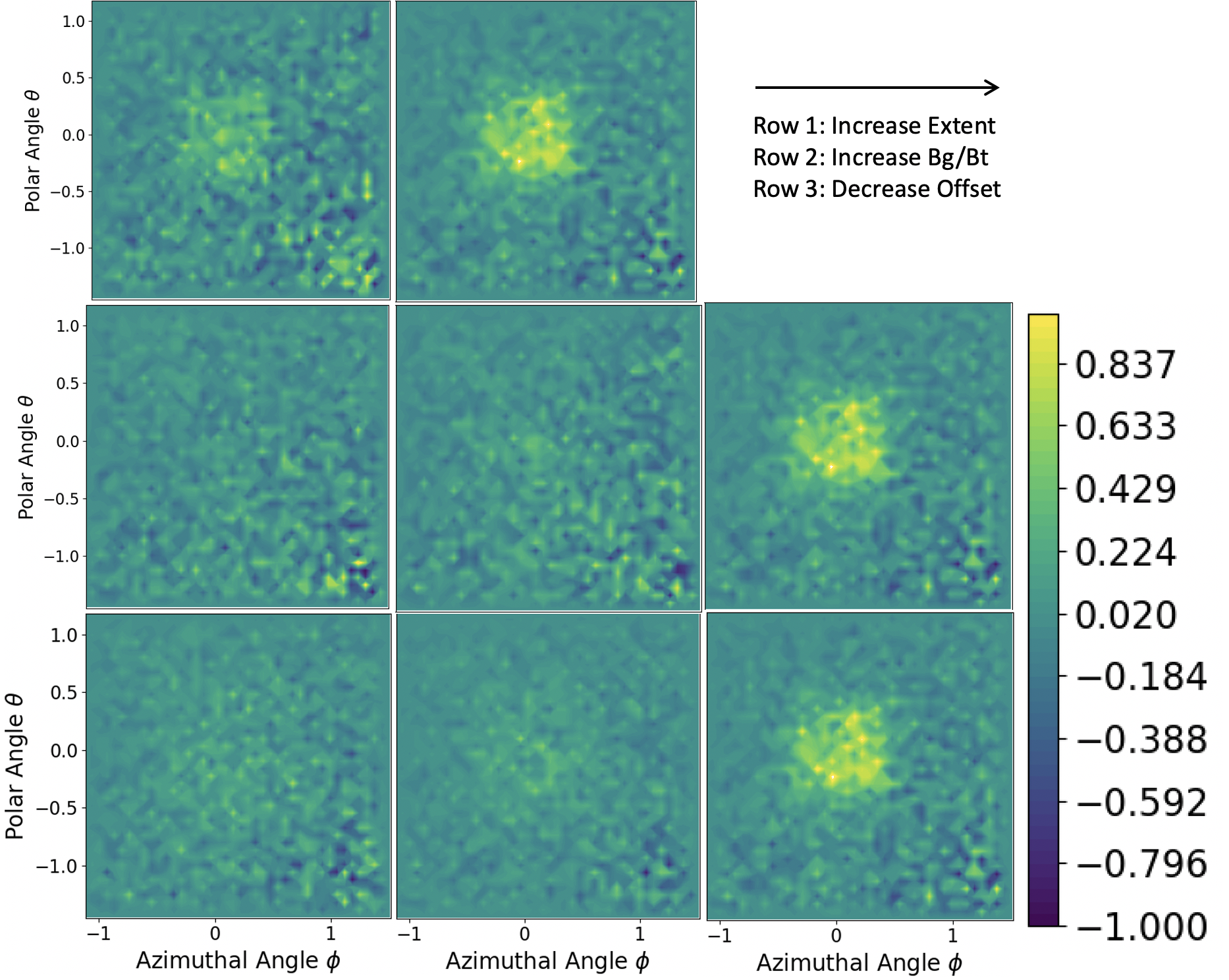}
\caption{Shows plots of $F(\Omega_z)/\sigma_{\Omega}$ for different box configurations using Method 2. The different rows show the effects on the Earth sky image of varying the box extent, coherent magnetic field ratio, and offset. The top row uses box extent values of 6 pc and 24 pc while keeping $B_{\text{g}}/B_{\text{t}}=1.0$ and offset=0 pc. The middle row uses values of $B_{\text{g}}/B_{\text{t}}=0.1$, 0.5, and 1.0 while keeping the box extent=24 pc and box offset=0 pc. The bottom row uses box offset values of 2 pc, 1 pc, and 0 pc while keeping the box extent=24 pc and $B_{\text{g}}/B_{\text{t}}=1.0$.}
\label{fig:method2}
\end{figure*}

Finally, we calculate for each plot a generic noise level $\sigma_\Omega$, where $\sigma_\Omega=|\text{min}(F(\Omega_z))|$. This simply means that the noise level for all of the angular bins for one plot is the magnitude of the most negative value computed, or the flux in the angular bin where the isotropic run had the highest flux above the main run. Though the uncertainty is technically a function of angular zone, this method is a simple way to calculate the uncertainty, and is an overestimate for most of the grid. Plotting $F(\Omega_z)/\sigma_\Omega$ allows us to neglect the proportionality constant in these expressions, as it should be the same for $F(\Omega_z)$ and $\sigma_\Omega$. The results are presented in Figures \ref{fig:method1} and \ref{fig:method2}, which show contour plots of $F(\Omega_z)/\sigma_\Omega$ at the earth point for Methods 1 and 2.

Each row in the figure shows the effect of varying one of the magnetic field configuration parameters; the box extent, coherent magnetic field ratio, and earth sky offset.  In these plots, the colored region in the upper left corner shows the location of the coherent magnetic field box. The effect of varying the three magnetic field configuration parameters produced what might be the expected changes on the background flux at the Earth sky. Increasing the box extent, decreasing its distance from the sky, and increasing the coherent magnetic field ratio produce a higher particle flux observed on the sky. This is in contrast to \cite{2016ApJ...822..102H}, where all coherent magnetic field values produce the same result as there is no turbulent field in the box to drive the particles out.We have varied the observer position and the magnitude of the anisotropy varies, but not significantly. 

The anisotropies presented here are much higher than the observed anisotropy in the cosmic ray flux, implying that the strength of the coherent magnetic fields can be much lower than what we assumed.  Even so, we are assuming the magnetic field energy in the coherent fields is much less than that of the small-scale structures.  The strength of our coherent magnetic fields was not chosen to match the observed anisotropies, but to demonstrate the role such magnetic fields can play on the flux observed at the earth.  Statistical limitations of our Monte Carlo method required higher coherent magnetic field strengths than those needed to explain the observations. Finally, we note that limitations with our grid setup, such as particles not being allowed to scatter back across the tally plane, may also have a minor effect on our results.

\section{Discussion and Conclusions}
\label{sec:conc}

In this study , we have extended the work of \cite{2016ApJ...822..102H} to include more realistic magnetic field structures which included both an isotropic and an anisotropic component. Additionally, we varied our magnetic field structures in a different way than was previously studied, varying the length of the field structure and moving it away from the tally plane. To accomplish this hybrid solution, we propose two methods which "combine" the standard transport and diffusion algorithms usually used. One of these methods involved the particles switching off between the two methods, picking the method of travel based on a probabilistic approach. The other method involved the addition of two vectors intended to represent the two components of the magnetic field separately.

We have shown that the anisotropies in the particle flux at the tally plane are much easier to create than might be expected, no matter which method is used. In particular, with a strong enough coherent magnetic field component, there are noticeable anisotropies above the background for all coherent field length scales and offset values tested. Because the length scale of the coherent field is so much larger than the mean free path of particle motion, even a weak coherent magnetic field component can give the particle a strong enough directional push over time to move it into the box region at the tally plane. Additionally, even magnetic field structures removed from the  plane can cause a noticeable particle flux above the background. This is because the coherent magnetic field alters the particle transport, effectively creating what appears to be a new cosmic ray source.  

Finally, we took the results at the tally plane and used them to make a construction of what an observer at Earth might see. Though this method had many limitations, we showed that the stronger, closer magnetic field configurations produce a noticeable anisotropy. Many other works have found a similar effect to the results presented here and in \cite{2016ApJ...822..102H}. In several studies  (\cite{2012PhRvL.109g1101G},\cite{2015ApJ...815L...2A}, \cite{2014PhRvL.112b1101A}, \cite{2016ApJ...830...19L}), it was shown that anisotropies in the flux can arise from local turbulent magnetic field structures.  Our simulations were only sensitive to anisotropies that are much larger than those observed.  The fact that modest coherent magnetic fields can produce strong anisotropies show that the features needed to explain the observed fields can be quite small and the small amplitudes of the observed cosmic ray anisotropies place limits on the nature of these coherent magnetic fields.

We stress that although we have made improvements over previous studies modeling cosmic ray transport, we have not done the ``ultimate" treatment of the fields. This involves resolving the turbulent magnetic field structure and using this in combination with the coherent field, directly solving the Lorentz equation of motion for every particle step. Although such full treatments are beyond computational power for large grids, it is possible to use small-scale calculations to improve on the recipes (Methods 1 and 2) used here.  We defer this more detailed study for a later paper.

Another way to reduce the uncertainties in this study would be to obtain better measurements of the magnetic field structures in the solar neighborhood.  With these structures, the cosmic ray anisotropies may be better understood.

\section*{Data Availability}

The data underlying this article are available in the article.

\section*{Acknowledgements}

This work was supported by the US Department of Energy through the Los Alamos National Laboratory. Los Alamos National Laboratory is operated by Triad National Security, LLC, for the National Nuclear Security Administration of U.S.\ Department of Energy (Contract No.\ 89233218CNA000001) This research was supported in part by the National Science Foundation under Grant No. NSF PHY-1748958 and by NASA ATP grant 80NSSC20K0507.

%%%%%%%%%%%%%%%%%%%%%%%%%%%%%%%%%%%%%%%%%%%%%%%%%%



\bibliographystyle{mnras}
\bibliography{main} % if your bibtex file is called example.bib

%\begin{thebibliography}{99}
%\end{thebibliography}

%%%%%%%%%%%%%%%%%%%%%%%%%%%%%%%%%%%%%%%%%%%%%%%%%%

%%%%%%%%%%%%%%%%% APPENDICES %%%%%%%%%%%%%%%%%%%%%

%%%%%%%%%%%%%%%%%%%%%%%%%%%%%%%%%%%%%%%%%%%%%%%%%%

% Don't change these lines
\bsp	% typesetting comment
\label{lastpage}
\end{document}